\documentclass[prd, aps, showpacs]{revtex4}
\usepackage[latin1]{inputenc} 

\usepackage[spanish,activeacute]{babel} 
\usepackage{graphicx}

\def\be{\begin{equation}}
\def\te{\end{equation}}
\def\ee{\end{equation}}
\def\ba{\begin{eqnarray}}
\def\bea{\begin{eqnarray}}
\def\nn{\nonumber\\}
\def\tea{\end{eqnarray}}
\def\ea{\end{eqnarray}}
\def\eea{\end{eqnarray}}

\begin{document}

\title{Real relativistic fluids in heavy ion collisions}

\author{E. Calzetta}
\email{calzetta@df.uba.ar}
\affiliation{Departamento de F\'\i sica, Facultad de Ciencias Exactas y Naturales, Universidad de Buenos Aires and IFIBA, CONICET, Ciudad Universitaria, Buenos Aires 1428, Argentina}

\begin{abstract}
The theory of real relativistic fluids is in the rather unique
situation that there is a natural relativistic extension of the
nonrelativistic theory, but it is physically untenable \cite{HisLin83}. On the
other hand, mounting evidence that matter created in relativistic
heavy ion collisions behaves as a relativistic fluid with small but
finite viscosity has given the quest for an alternative a definite
goal \cite{Risc98}. We shall review different approaches to relativistic real
fluids, their link to relativistic kinetic theory, and their
application to the analysis of heavy ion collisions  \cite{CalHu08} 
\end{abstract}
\maketitle

\section{Introduction} 
The goal of these lectures is to discuss the theory of relativistic real fluids under the excuse of its application to the description of relativistic heavy ion collisions \cite{Roma09,Ruus86}. More concretely, our goal is to discuss, in the simplest possible terms, why the ``Einstein elevator'' concept is not useful in the development of this theory. Namely, we cannot build a theory of relativistic real fluids by just asking that it reduces to its nonrelativistic counterpart, namely the Navier - Stokes equations, in an inertial frame where the fluid is at rest. For this reason we shall begin with a ``derivation'' of the nonrelativistic theory of ideal fluids (those which  flow with no entropy production) from thermodynamics \cite{ChaLub95}. For ideal fluids, the relativistic theory may be found as the covariant extension of the nonrelativistic theory enforced in the rest frame. We shall then show that already in this simple framework we may obtain a working picture of how a relativistic heavy ion collision works, yielding remarkably accurate predictions regarding several concrete observables. 

Among these observables there is the so-called elliptic flow, namely the anisotropy of the particle yield in the transverse plane to the beam direction. If the matter deposited in the collision region behaves as a nearly ideal fluid, then elliptic flow arises as the simple transduction of pressure gradients in the original configuration into velocities in the asymptotic expanding state. This is an obvious prediction of the Euler equations which is surprisingly hard to reproduce in non-hydrodynamic models. The success in predicting that there must be an elliptic flow is one of the most compelling reasons to believe that matter in the collision region, from some very early time after the actual collision up to the break up in the individual hadrons which are eventually detected, behaves as a fluid. However, ideal hydrodynamics overestimates the amount of elliptic flow \cite{Roma09}. This suggests there must be a mechanism that tends to isotropize flow, and indeed viscosity acts in precisely that direction. Therefore, the next natural step is to consider hydrodynamic models based on real fluid hydrodynamics.

At this point we return to the formal theory, and generalize our previous ``thermodynamic'' derivation to the case of real fluids. The resulting non relativistic theory is given by the Navier - Stokes equations, which is satisfactory, but the relativistic theory emerging from the ``Einstein elevator'' is essentially flawed - if covariance is enforced, then the theory has no stable solutions \cite{HisLin83}. How to get out of this cul de sac is the subject of the rest of the lectures.

Since the ``thermodynamic'' approach led us nowhere, we shall begin anew from a more fundamental point of view, that provided by relativistic kinetic theory. In this case, there is an agreed upon relativistic formulation to build on \cite{Isr72,Isr88}. The problem is how to reduce it to the hydrodynamic level. In the nonrelativistic case, the method of choice is the so-called Chapman - Enskog expansion, which leads to the Navier - Stokes equations \cite{CHACO39}. In the relativistic case, it works likewise, and so it must be rejected. However, a second well known technique, the Grad expansion, is successful in building a relativistic hydrodynamics where linear perturbations of an equilibrium state evolve causally. 

However, the application to relativistic heavy ion collisions forces us to face the relationship between kinetic and hydro descriptions a second time, because what is actually measured are individual hadrons. This means that there must be some space time ``break up'' or ``freeze out'' surface (or region) where the fluid nucleates into individual particles that fly away to the detectors (a process that is by no means simple, but does not belong to these lectures). We must be able to read the one particle distribution function of these particles out of the hydrodynamic state of the fluid just before break up. If we simple invert the Grad expansion, we are led to a one particle distribution function which is not nonnegative throughout phase space, a defect that contaminates the predictions of the theory regarding observables which are sensitive to high momenta.

This drawback of the Grad approach, and also the need to include nonlinear corrections to the linearized equations of motion without spoiling stability and covariance, prompted us to suggest that the Grad approach may be just the linear approximation to a more comprehensive view of the kinetic - hydro relationship \cite{CalPR10}. The basic insight is that any realistic kinetic theory displays a whole hierarchy of relaxation times, from the relaxation times characteristic of large scale inhomogeneities to the much shorter relaxation times of hard modes. Fluctuations in the hydrodynamic modes relax on the longer time scales, and are perceived by the harder modes as externally imposed thermodynamic forces which prevent their relaxation to true equilibrium. In other words, the truly kinetic modes relax not to equilibrium but to a nonequilibrium steady state constrained by the instantaneous configuration of the hydrodynamic modes. It has been known from long ago that such steady states are the solutions to variational problems \cite{Ono61,Jay80}. Prigogine among others has proposed that they are the extrema of the entropy \emph{production}, as opposed to the extrema of the entropy itself, which are the true equilibria \cite{Pri55}. Known proofs of the so-called ``Prigogine theorem'' are restricted to linear irreversible thermodynamics \cite{Lan75,Bru06}; we shall appeal to it on a heuristic rather than formal basis - the idea is that most kinetic modes are in the linear regime most of the time anyway, so a theory which is good in the linear regime is good enough to compute global observables such as stress tensor components, but we shall  not attempt to formulate this insight in any rigorous way \cite{Kli91,Kli95}.

We shall therefore conclude these lectures with a brief presentation of this ``entropy production variational method'', its relationship to positivity of the one particle distribution function, and its nonlinear generalization.

The lectures are organized as follows. In next section we present the theory of relativistic ideal fluids, as derived from thermodynamics, and its application to the description of relativistic heavy ion collisions (RHICs). The main success of the theory, namely the prediction of elliptic flow, is also its fatal drawback, because experimental data show that ideal hydrodynamics overestimates the flow anisotropy \cite{Roma09}. We are therefore motivated to extend the theory to include viscosity, whereby we hit on the stability problem \cite{HisLin83,Ols90,OlsHis90}. We present the theory of real fluids in Section III; since the thermodynamic derivation fails, we base our discussion on relativistic kinetic theory. We must confront the \emph{closure} problem, that is, to relate the one particle distribution function in the kinetic description to the hydrodynamic degrees of freedom. We present the Chapman-Enskog solution (which reproduces the results from the thermodynamic analysis, and therefore is unsuitable) and the Grad solution, and finally identify the Grad solution of the closure problem as the linearized approximation to an approach based on the entropy production variational method (EPVM). We conclude with some brief final remarks.

If relativistic hydrodynamics may be founded on kinetic theory, kinetic theory itself comes from quantum field theory, the main subject matter of this school. We show the main lines of the derivation of kinetic theory from quantum field theory in the Appendix, which relies heavily on ref. \cite{CalHu08}.

\section{From Thermodynamics to Hydrodynamics}
\subsection{Nonrelativistic Ideal fluids}
Let us begin by reviewing non relativistic hydrodynamics, as ``derived'' from thermodynamics.
The basic tenets of thermodynamics we need to keep in mind are the following: we
have a (simple) system described by some intensive parameters (temperature $T$,
chemical potential $\mu$, pressure $p$, etc.) whose meaning we take for granted (e.g.
we already know everything about the zeroth law) and extensive parameters,
such as energy $U$, entropy $S$, volume $V$ , and particle number $N$ (we use particle
number for concreteness, but any, or several, conserved charge(s) would serve
just as well). In equilibrium, all these quantities are position independent. Their
first deviations from equilibrium are related by the first law

\be
TdS = dU + pdV -\mu dN 
\label{(12.1)}
\te
Extensive quantities are homogeneous functions of each other, so we must have

\be
TS = U + pV - \mu N 
\label{(12.2)}
\te
From the differential of this second identity we obtain the Gibbs-Duhem relation

\be
dp = sdT + nd\mu 
\label{(12.3)}
\te
where $s = S/V$ and $n = N/V$ are the entropy and particle number densities,
respectively. This means

\bea
\left.\frac{\partial p}{\partial T}\right|_{\mu}
&=& s = \frac{\rho + p - \mu n}T\nn
\left.\frac{\partial p}{\partial \mu}\right|_T
&=& n 
\label{(12.4)}
\tea
where $\rho = U/V$ is the energy density.

Let us now consider a non equilibrium, inhomogeneous configuration. The above relations still hold for the densities of the conserved quantities energy, momentum (which we must include) and particle number. The dynamics is given by the conservation laws

\bea
\frac{\partial \rho}{\partial t}&=&-\nabla_j J^ j_u\nn
\frac{\partial p^ i}{\partial t}&=&-\nabla_j T^{ij}\nn
\frac{\partial n}{\partial t}&=&-\nabla_j J^j
\tea 
Here $\rho$, $p^i$ and $n$ are the energy, momentum and charge densities, $J^j_u$, $T^{ij}$ and $J^j$ the respective currents. $T^{ij}$ must be symmetric \cite{Bat67}. The entropy density

\be 
s=\frac 1T\left[ \rho+p-v_ip^i-\mu n\right] 
\te 
obeys the first Law

\be 
Tds=d\rho-v_idp^i-\mu dn
\te 
which implies the Gibbs - Duhem relation

\be 
dp=sdT+p^idv_i+nd\mu 
\te 
In these formulae we have introduced the velocity $v_i$ and the chemical potential $\mu $. We therefore have

\bea 
\frac{\partial s}{\partial t}&=& \frac 1T\left\lbrace \frac{\partial \rho}{\partial t}-v_i\frac{\partial p^ i}{\partial t}-\mu\frac{\partial n}{\partial t}\right\rbrace \nn
&=& \frac {-1}T\left\lbrace \nabla_j J^ j_u-v_i\nabla_j T^{ij} -\mu\nabla_j J^j\right\rbrace 
\tea 
We then have

\be 
\frac{\partial s}{\partial t}=-\nabla_j\left\lbrace\frac 1T\left[J^ j_u-v_iT^{ij}-\mu J^j \right]  \right\rbrace +\left[J^ j_u-v_iT^{ij}-\mu J^j \right]\nabla_j\left(\frac 1T \right) -\frac 1TT^{ij}\nabla_jv_i-\frac {1}TJ^j\nabla_j\mu 
\te
To proceed we write $J^j=nv^j+\chi^j$. In the terms involving the velocity we eliminate the  gradient of the chemical potential by using the Gibbs - Duhem relation 

\bea 
\frac{\partial s}{\partial t}&=&-\nabla_j\left\lbrace\frac 1T\left[J^ j_u-v_iT^{ij}-\mu J^j \right]  \right\rbrace +\left[J^ j_u-v_iT^{ij}-\mu J^j \right]\nabla_j\left(\frac 1T \right) -\frac 1TT^{ij}\nabla_jv_i\nn
&-&\frac 1T{v}^j\left[\nabla_jp-s\nabla_jT-p^i\nabla_jv_i \right] - \frac {1}T\chi^j\nabla_j\mu  
\tea
so far

\bea 
\frac{\partial s}{\partial t}&=&-\nabla_j\left\lbrace\frac 1T\left[J^ j_u+p{v}^j-v_iT^{ij}-\mu J^j \right]  \right\rbrace \nn
&+&\left[J^ j_u+p{v}^j-v_iT^{ij}-\mu J^j-Ts{v}^j \right]\nabla_j\left(\frac 1T \right) \nn
&-&\frac 1T\left[ T^{ij}-p^i{v}^j-p\delta^{ij}\right] \nabla_jv_i- \frac {1}T\chi^j\nabla_j\mu
\tea
Using again the second law, we may write

\be
J^ j_u+p{v}^j-v_iT^{ij}-\mu J^j-Ts{v}^j =J^ j_u-\left(\rho+p\right)v^j-v_i\left(T^{ij}-p^iv^j-p\delta^{ij}\right)-\mu\chi^j 
\te
and so

\bea 
\frac{\partial s}{\partial t}&=&-\nabla_j\left\lbrace\frac 1T\left[J^ j_u+p{v}^j-v_iT^{ij}-\mu J^j \right]  \right\rbrace \nn
&+&\left[J^ j_u-\left(\rho+p\right)v^j-v_i\left(T^{ij}-p^iv^j-p\delta^{ij}\right)\right]\nabla_j\left(\frac 1T \right) \nn
&-&\frac 1T\left[ T^{ij}-p^i{v}^j-p\delta^{ij}\right] \nabla_jv_i-\chi^j\nabla_j\alpha
\label{dsdt}
\tea
Where $\alpha=\mu /T$. Since we have allowed for the possibility of the energy and particle number flux not being collinear, we must adopt a convention about what is the fluid velocity. We shall adopt the so-called Landau-Lifshitz prescription, namely, that $J_u^j$ and $p^j$ vanish when $v^j=0$ \cite{LL6}.

We define an ideal fluid as one that flows with no entropy production. Thus for an ideal fluid

\bea
T^{ij}&=&p^i{v}^j+p\delta^{ij}\nn
J^ j_u&=&\left(\rho+p\right)v^j\nn
\chi^j&=&0
\tea
The symmetry of $T^{ij}$ requires $p^i$ to be proportional to $v^i$. The equations for an ideal fluid are thus

\bea
\frac{\partial n}{\partial t}+\nabla_j nv^j&=&0\nn
\frac{\partial \rho}{\partial t}+\nabla_j \left(\rho+p\right)v^j&=&0\nn
\frac{\partial p^ i}{\partial t}+\nabla_j p^iv^j+\delta^{ij}\nabla_jp&=&0
\tea 

\subsection{Relativistic Ideal Fluids}
We now generalize the above framework of thermodynamics to a relativistic fluid
evolving in a spacetime with an arbitrary metric $g_{\mu\nu}$. All derivatives shall be covariant derivatives
with respect to the Levi-Civita connection, so that $g_{\mu\nu;\rho}=0$. We write the Minkowsky metric as $\eta^{00}=-1$, $\eta^{0i}=0$ and $\eta^{ij}=\delta^{ij}$, where a $0$ refers to coordinate $x^0=ct$

To simplify matters we will describe the construction of a covariant theory in
terms of a set of rules \cite{Isr88}:

(a) Intensive quantities ($T$, $p$, $\mu$) are associated with scalars, which represent the
value of the quantity at a given event, as measured by an observer at rest
with respect to the fluid.

(b) Extensive quantities ($S$, $V$, $N$) are associated with vector currents $S^{\mu}$, $u^{\mu}$, $N^{\mu}$. If a given observer measures a density $x$ and a flux $J_x^i$ for the quantity $X$, then $X ^{\mu}=\left(x,J_x^i/c\right)$ The quantity $u^{\mu}$ associated with volume is the fluid
4-velocity; in Minkowsky space

\be
u^{\mu}=\frac1{\sqrt{1-\frac{v^2}{c^2}}}\left(1,\frac{v^i}c\right)
\te
$u^{\mu}$ obeys the additional constraint $u^2 = -1$. We call density tout
court the density measured by an observer comoving with the fluid, namely
$x = -u_{\mu}X^{\mu}$.

If the quantity $X$ is conserved,
then $X^{\mu}_{;\mu} = 0$. If we consider the variation of the entropy content within some spatial region 
as a function of time, the second law demands that the increase in entropy should
be higher than the entropy flow through the boundary . Thus the covariant
statement of the second law is that entropy production must be positive, i.e.
$S^{\mu}_{;\mu}\ge 0$.

(c) Energy and momentum are combined into a single extensive quantity
described by an energy-momentum tensor $T^{\mu\nu}$ which is symmetric. Therefore any given observer will identify the energy density as $T^{00}$, the energy flux as $cT^{0i}$, the momentum density as $T^{i0}/c$ and the momentum flux as $T^{ij}$. The symmetry of the energy momentum tensor implies that the momentum density is $c^{-2}$ times the energy flux \cite{LL6}. We define the rest frame of the fluid as the frame where the energy flux vanishes. As before, the energy density $\rho$ is defined as the energy density in the rest frame. Both the fluid velocity and energy density may be found from the eigenvalue equation $T^{\mu\nu}u_{\nu}=-\rho u^{\mu}$

Let us now describe a relativistic ideal fluid, namely one that flows with no entropy production. We shall proceed by correspondence between the relativistic and non relativistic theories. To this end, we assume that in the rest frame we have the decomposition

\be
N^{\mu}=\left(n,0\right)
\te

\be
T^{\mu\nu}=\left(
\begin{array}{cc}
\rho&0\\
0&p\delta^{ij}
\end{array}\right)
\te
So in an arbitrary frame we must have

\bea
N^{\mu} &=& nu^{\mu}\nn
T^{\mu\nu}&=& \rho u^{\mu}u^{\nu}+ p\Delta^{\mu\nu}
\label{(12.11)}
\tea
where $\Delta^{\mu\nu}=g^{\mu\nu}+u^{\mu}u^{\nu}$. The entropy
current $S^{\mu}$ in the rest frame is given by $S^{\mu}=\left(s,0\right)$, where $Ts = p + \rho - \mu n$. In an arbitrary frame $S^{\mu}=su^{\mu}$ or

\be
S^{\mu}= -T^{\mu\nu}\beta_{\nu} + pu^{\mu} - {\alpha}N^{\mu}
\te
Where the inverse temperature vector $\beta^{\mu} = T^{-1}u^{\mu}$ and $\alpha=\mu /T$. Observe that $T^{-2} = -\beta_{\mu}\beta^{\mu}$. Then 

\be
S^{\mu}_{;\mu} = -\beta_{\nu}T^{\mu\nu}_{;\mu}- \alpha N^{\mu}_{;\mu} 
\label{ (12.15)}
\te
This means that entropy production vanishes for an ideal fluid, provided the conservation
laws of energy-momentum and particle number hold.

\subsection{Ideal Hydrodynamic models of heavy ion collisions}

The theory of relativistic real fluids has been brought to the limelight by the mounting evidence that one such system has been seen in actual experiments, namely, relativistic heavy ion collisions (RHICs). Indeed, one can get a working understanding of RHICs from the theory of ideal fluids we have just described. Nevertheless, precisely the demands of a better fit between theory and experiment will prompt us to develop a theory or real fluids, namely, to add viscosity to the flow equations. In this section we shall briefly describe RHICs, and discuss what can and cannot be reproduced by models based on ideal hydrodynamics.

RHICs have been produced in several experiments, namely the Super Proton Synchrotron (SPS) at CERN, the Relativistic Heavy-Ion Collider (RHIC) at Brookhaven and
the Large Hadron Collider (LHC) once again at CERN. The
RHIC experiments in particular are described in detail in the so-called `white
papers', which are possibly the most reliable source on the subject \cite{BRAHMS05,PHOBOS05,STAR05,PHENIX05}. 
One of the goals of the RHIC program is to probe into possible new
phases of nuclear matter at high energies.  In such a high energy phase, matter is expected to form a plasma of gluons and (massless) quarks (quark-gluon
plasma, QGP).

Virtually all theoretical analysis of RHICs assume a
space-time picture of the collision provided by the Bjorken
model \cite{BJOR83}. The colliding nuclei are
seen as slabs of quark and gluon matter. In the center of mass frame,
both slabs approach each other at near light speed. Upon collision,
the two slabs of matter will mostly go through each other, leaving
behind a wake of hot plasma. We may then distinguish three different
regions: the two fragmentation regions
corresponding to the receding slabs, and the central
region corresponding to the plasma in between.
We are interested in phenomena in the central region.

At the time of crossing a number of hard scattering processes will
occur, whose products will reach directly the detectors. These hard
processes are unrelated to the nonequilibrium dynamics of the plasma;
and may presumably be predicted on perturbative QCD grounds. In what
follows, we will assume this hard component has been isolated despite
great difficulty to achieve this in reality.

The hot plasma will expand and cool, and eventually fragment into
ordinary particles in flight intercepted by the detectors. We wish to
predict the number of particles of each species to be detected, as a
function of the angle $\theta $ between the direction of flight and
the direction $z$ of the beam. It is remarkable that with this simple
picture we can state a first observable prediction already.

Indeed, because of Lorentz contraction, we may think of the
approaching slabs as infinitely thin in the direction of motion $z$,
and in a first approach to the problem, as infinite and homogeneous
in the transverse directions $x$ and $y$. This picture is invariant
under boost in the $z$ direction, and so is the final
distribution of particles. So if we parametrize the momentum of an
out-going particle as $p^{0}=E$, $p^{3}=p$ and $\left(
p^{1},p^{2}\right) =p_{\perp }$, then the distribution of
particles may depend only upon the transverse momentum and $E^{2}-p^{2}=m^{2}+p_{\perp }^{2}$. In particular, it must be independent of $\theta ,$ since $\cos \theta \sim p/E$ is not invariant. It is
conventional to plot the yield of the collision in terms of the
\textit{rapidity} $Y$, defined by $p/E\equiv \tanh
Y$, or rather the pseudorapidity $\eta =-\ln
\tan \left[ \theta /2\right] ,$ $\tanh \eta =p/\left|
\mathbf{p}\right| .$ Rapidity and pseudo-rapidity agree at momenta
which are large compared to the mass of the particle. Then the
prediction in this picture is that there is a \textit{plateau} in the
(pseudo) rapidity distribution, at least for
small rapidity ($\left| \eta \right| \rightarrow \infty $ corresponds
to the fragmentation rather than the central region). Although this
prediction is not quantitatively borne out by the RHIC data \cite{PHOBOS05}, the experimental curve flattens enough at low rapidity that it may be accepted as a working rough approximation.

We may elaborate the Bjorken picture further. Let us assume that the
plasma is formed on the plane $z=0$ at the time $t=0$ of the
collision, and then expands along the $z$ direction. A given plasma
element will cool according to its own proper time $\tau $. 
Eventually, at some given constant $\tau $ surface, the plasma will
be cold enough (and/or dilute enough) to break up into hadrons.
Assuming that the product hadrons are thermally distributed,
massless and at zero chemical potential, the Bose-Einstein
distribution predicts that the energy per particle is $\epsilon /n=2.7$
$T$. Since temperature is constant on the break up surface, this
means that in all collisions particles should have the same average
energy. Indeed, it is observed that the energy per particle is about
$0.8$ \textrm{GeV}, regardless of center-of-mass energy and impact
parameter.

To obtain a more quantitative description of the process, we 
describe the plasma as a relativistic ideal fluid.  To
close the hydrodynamic system of equations we must provide the
equation of state. The central feature of this is the
``softening'' near the critical point, meaning that the speed of sound $%
c_{s}^{2}=\partial p/\partial \epsilon \rightarrow 0$ as we approach the
transition point. The softening of the equation of state affects the
evolution of the fireball, which then becomes a signal of whether the
transition point has been reached or not.

Since perfect fluids conserve entropy, the total entropy within the
fireball remains constant, and $T$ scales as $V^{-1/3}.$ So, if the
expansion is one-dimensional, and we consider the volume enclosed
between two fixed rapidities, then $T\sim \tau ^{-1/3},$ where $\tau
$ is the proper time. In particular, the energy density scales as
$\tau ^{-4/3}$.

We now consider more closely the phenomenon of
break-up \cite{Roma09,Ruus86}. Assume this occurs on a
3-dimensional surface $\Sigma$ defined by some equation $\Sigma
\left( x^{\mu }\right) =0.$ If $x_{0}$ is a solution, then the normal
vector at $x_{0}$ is $n^{\mu }=\left( \alpha \right) \Sigma^{,\mu
}$, $\alpha =\left( -\Sigma _{,\mu }\Sigma ^{,\mu }\right) ^{-1/2}.$
We shall assume that $n^{\mu }$ is timelike. The invariant measure on $\Sigma $ is given by
$d^{3}\sigma =d^{4}x\;\delta \left( \Sigma \right) \alpha ^{-1}.$

Let us assume that right up to break-up we can describe
matter as a perfect relativistic fluid, and as a noninteracting relativistic gas thereafter. Let $K_{a}=\partial /\partial
x^{a}$ be the four Killing vectors of Minkowski space. Then Gauss'
theorem shows that the quantities $n_{\mu }K_{a\nu }T^{\mu \nu }$ and
$n_{\mu }N^{\mu }$ are continuous across the break-up surface (we
shall consider only one conserved current, corresponding to, say, the
baryon number). These conditions plus the equation of state of the
hadronic phase define the energy density, pressure, baryon number
density (or equivalently, the temperature and chemical potential) and
the four-velocity of the hadrons at break-up. The detailed spectrum
is found by assuming that the hadrons are thermally distributed.

The total number of emitted particles is

\begin{equation}
\int d^{3}{\mathbf {x}}\;K_{0\mu }N_{had}^{\mu }
\end{equation}
where the integral is over some $t=$\textrm{cons}$\tan $\textrm{t}
surface well to the future of the collision. Because of Gauss
theorem, we may replace the integral by an integral over the
break-up surface (we may have to complete this surface to get a
Cauchy surface, but the particle density flux will vanish on these
additions anyway). But then we may use the matching conditions to
express this integral in terms of the particle current before
break-up. We obtain the total number of emitted particles as

\begin{equation}
\int d^{4}x\;\delta \left( \Sigma \right) \Sigma _{,\mu }N_{hydro}^{\mu }
\end{equation}
In practice, we may wish to smear a little the position of the
break-up surface, thus writing the total number of emitted particles
as

\begin{equation}
\int d^{4}x\;\left[ \frac{e^{-\Sigma ^{2}/2\left( \Delta \Sigma \right) ^{2}}%
}{\sqrt{2\pi }\left( \Delta \Sigma \right) }\right] \Sigma _{,\mu
}N_{hydro}^{\mu }
\end{equation}

The total number of particles of species $i$ with momentum $p^{\mu }$ is

\begin{equation}
g_{i}\int {d^{4}x}\;\left[ \frac{e^{-\Sigma
^{2}/2\left( \Delta \Sigma \right) ^{2}}}{\sqrt{2\pi }\left( \Delta \Sigma
\right) }\right] \delta \left(
p_{i}^{2}-m_{i}^{2}\right) \frac{\Sigma _{,\mu }p_{i}^{\mu }}{\left[
\exp \left( -\beta _{\nu }p_{i}^{\nu }-\mu_{i}\right) -\varepsilon _{i}%
\right] }  \label{spec1}
\end{equation}
where the temperature four vector and chemical potentials are read out from the (ideal) energy momentum tensor and charged currents of the fluid just before break up.

The two basic observables are the total number of particles with
transverse (respect to the beam axis) momentum $p_{\perp }$, which is
usually given in terms of the transverse mass
$m_{\perp }^{2}=m^{2}+p_{\perp }^{2},$ and the elliptic
flow coefficient $v_{2},$ which results from
fitting the particle spectrum in the transverse plane to a second
harmonic $\left( 1+2v_{2}\left( p_{\perp }\right) \cos 2\phi \right)
,$ where $\phi $ is the angle measured from the reaction plane. This
is equivalent to considering an elliptic fireball, in which case
$v_{2}$ measures the eccentricity of the ellipse. The first harmonic
is called directed flow, and would represent a
shifted spherical fireball in the transverse plane.

The agreement of predictions from hydrodynamical simulations with
experimental data is good, provided the simulation is started very
early (earlier than $1$ \textrm{fm}/$c$ after the collision). If one
believes that the validity of hydrodynamics demands (local)
equilibration, this very short time is somewhat of a puzzle. 

\subsection{Elliptic flow}

Of the several predictions of hydrodynamic models, elliptic flow is one of the most compelling, because it is hard to match by alternative models. Let us show how elliptic flow arises in hydro models. 
 
Let the velocity of the fluid (with respect to laboratory time) in the longitudinal direction be $dz/dt=v^z$. We may define a longitudinal rapidity

\be 
y_L=\frac12\ln\frac{1+v^z}{1-v^z}
\te 
Let us take a boost invariant flow in the beam direction as background solution. Then $v^z=z/t$ and the longitudinal rapidity becomes identical with the space time rapidity

\be 
y_L=\eta=\frac12\ln\frac{t+z}{t-z}
\te 
We introduce Milne time $\tau=\sqrt{t^2-z^2}$ and Milne coordinates $\xi^{\alpha}=\left( \tau,\eta,x,y\right) $. Define
${dx^{\mu}}/{d\tau}= \left( t/\tau,z/\tau,\vec{v}_{\perp}\right) $. We find 

\be 
-\eta_{\mu\nu}\frac{dx^{\mu}}{d\tau}\frac{dx^{\nu}}{d\tau}=1-{v}_{\perp}^2
\te 
Introducing the transverse rapidity 

\be
y_{\perp}=\frac12\ln\frac{1+{v}_{\perp}}{1-{v}_{\perp}}
\te 
then $v_{\perp}=\tanh y_{\perp}$ and the four velocity becomes

\be 
u^{\mu}=\cosh y_{\perp}\;\left( \cosh\eta,\sinh\eta,\vec{v}_{\perp}\right)
\te 
or else in Milne coordinates

\be 
u^{\alpha}=\cosh y_{\perp}\;\left( 1,0,\vec{v}_{\perp}\right)
\te 
The interval element in Milne coordinates is $ds^2=-d\tau^2+\tau^2d\eta^2+dx^2+dy^2$, and so there are nontrivial Christoffel symbols

\be 
\Gamma^{\tau}_{\eta\eta}=\tau
\te 
and

\be 
\Gamma^{\eta}_{\tau\eta}=\frac 1{\tau}
\te 
In the strictly boost invariant case $y_{\perp}=0$ we have $u_{\eta ;\eta}=\tau$ and $u^{\alpha}_{;\alpha}=1/\tau$, although $u^{\beta}u_{\alpha ;\beta}=\Gamma^{\tau}_{\alpha\tau}=0$. The covariant Euler equation is identically satisfied, $n\propto \tau^{-1}$, while the energy density $\rho\propto \tau^{-4/3}$ for conformal matter. 

Obviously boost invariant expansion cannot create elliptic flow. The next step is to consider a nontrivial dynamics in the transverse plane. 
It is convenient to parametrize the transverse plane in terms of Lagrangian coordinates $\vec{q}$. Then the full coordinates are $\zeta^{\rho}=\left( \tau,\eta,q^ 1,q^ 2\right) $; the transformation back to Euler coordinates is given by the system

\be 
 \frac{\partial x^i }{\partial \tau}\vert_{q^ a}= v^i\left( \tau, x^i\right) 
\te 
with initial condition $x^i\left( \tau =0\right) =q^ i$. 
Now write

\bea
ds^2&=&-d\tau^2+\tau^2d\xi^2+\delta_{ij}dx^idx^j\nn
&=&-d\tau^2+\tau^2d\xi^2+\delta_{ij}\left( \frac{\partial x^i}{\partial q^ a}dq^a+v^id\tau\right) \left( \frac{\partial x^j}{\partial q^ b}dq^b+v^jd\tau\right)
\tea
In the new coordinates the metric reads

\be
{g_{\rho\sigma}}=\left(
\begin{array}{ccc}
-\left( 1-g^{ab}N_aN_b\right)  & 0  & N_b
\\
0 & \tau^{2} & 0
\\
N_a & 0 & g_{ab}
\end{array}
\right)
\te
where

\be 
g_{ab}=\delta_{ij} \frac{\partial x^i}{\partial q^ a}\frac{\partial x^j}{\partial q^ b}
\te 

\be 
N_a=v_i\frac{\partial x^i}{\partial q^ a}
\te 
Observe that $g^{ab}N_aN_b=v_iv^i=\tanh^2y_{\perp}$, so $g_{00}=-1/\cosh^2y_{\perp}$. 
The lapse function is then $N=1$ and the four metric determinant is $g=- \tau^{2} g^{\left( 2\right) }$, where $g^{\left( 2\right) }=\mathrm{det}\;g^{ab}$. In Milne-Lagrange coordinates the four velocity is $u^{\alpha}=\left( \cosh y_{\perp},\vec{0}\right) $, so 

\be 
u^{\nu}_{;\nu}=\frac 1{\tau\sqrt{g^{\left( 2\right) }}}\frac{\partial}{\partial \tau}\tau\sqrt{g^{\left( 2\right) }}\cosh y_{\perp}
\te
and the ideal hydrodynamic equations

\bea
\frac{\dot{T}}T+\frac 13u^{\nu}_{,\nu}&=&0\nn
\dot{u}^{\mu}+\Delta^{\mu\nu}\frac{T_{,\nu}}T&=&0
\tea
Give

\be 
T=\frac{T_0\left(\vec{q} \right) }{\left[ \tau\sqrt{g^{\left( 2\right) }}\cosh y_{\perp}\right] ^{1/3}}
\te 
and

\be 
\cosh^2y_{\perp}\Gamma^a_{00}+\frac 1T\left[ N^aT_{,\tau}+g^{ab}T_{,b}-N^aN^bT_{,b}\right] =0
\te
We shall consider only linear deviations from homogeneity, where

\be 
x^a=q^a+\xi^a\left(\vec{q},\tau \right) 
\te
$\xi^a\left(\vec{q},0 \right) =0$
\be 
N^a=v^a=\frac{\partial}{\partial \tau}\xi^a\left(\vec{q},\tau \right) 
\te

\be
\sqrt{g^{\left( 2\right) }}=1+\xi^a_{,a}
\te

\be 
\Gamma^a_{00}=\frac{\partial^2}{\partial \tau^2}\xi^a\left(\vec{q},\tau \right) 
\te
So, writing $T_0\left(\vec{q} \right) =T_0\left(1+\delta \left(\vec{q} \right)\right)$

\be 
\frac{\partial^2\xi_a}{\partial \tau^2} -\frac 1{3\tau}\frac{\partial\xi_a}{\partial \tau}-\frac 1{3}\xi^b_{,ba} =-\delta_{,a}
\te
Given our initial conditions, there are no transverse velocity components. Writing $\xi_a=\phi_{,a}$ we get

\be 
\frac{\partial^2\phi}{\partial \tau^2} -\frac 1{3\tau}\frac{\partial\phi}{\partial \tau}-\frac 1{3}\bf{\Delta}\phi=-\delta
\te
Write

\be
\delta=\int\frac{d^2k}{\left(2\pi\right)^2}\;e^{ikq}\delta_k
\te
Then

\be
\phi=\int\frac{d^2k}{\left(2\pi\right)^2}\;e^{ikq}\phi_k
\te

\be 
\frac{\partial^2\phi_k}{\partial \tau^2} -\frac 1{3\tau}\frac{\partial\phi_k}{\partial \tau}+\frac {k^2}{3}\phi_k=-\delta_k
\te
Write $\dot{\phi_k}=\varphi_k$

\be 
\frac{\partial^2\varphi_k}{\partial \tau^2} -\frac 1{3\tau}\frac{\partial\varphi_k}{\partial \tau}+\frac13 {\left[k^2+\frac 1{\tau^2}\right]}\varphi_k=0
\te
The solution which is regular at $\tau=0$ is $\varphi_k=c_k\tau^{2/3}J_{1/3}\left(k\tau/\sqrt{3}\right)$. As $\tau\to 0$ we get $\varphi_k\approx c_k\tau\left(k/\sqrt{3}\right)^{1/3}$, $\phi_k\approx c_k\tau^2\left(k/\sqrt{3}\right)^{1/3}/2$, so

\be
c_k=\frac{-3}{2}\left(\frac k{\sqrt{3}}\right)^{-1/3}\delta_k
\te
We may now find the velocities

\be
v_a=\frac{\partial}{\partial \tau}\phi_{,a}=\varphi_{,a}=\left(\frac{-3^{3/2}}{2}\right)\int\frac{d^2k}{\left(2\pi\right)^2}\;e^{ikq}\frac{ik_a}k\left(\frac {k\tau}{\sqrt{3}}\right)^{2/3}\delta_kJ_{1/3}\left(k\tau/\sqrt{3}\right)
\te
At early times $v\propto -\tau\nabla\delta$, and thus the flow transforms an initial density (or pressure) gradient into a flow anisotropy. This simple effect is remarkably hard to reproduce in parton models, under any realistic initial condition.

The important conclusion is that even in this simple model we get a qualitative picture of a relativistic heavy ion collision that serves as a framework for further analysis. In reality the flow is not really boost invariant, and it is certainly not homogeneous in the transverse plane. The hydrodynamic models offer a natural way to couple flow to pressure gradients in the transverse plane, and in this respect are superior to all other known alternatives. On the other hand, ideal hydrodynamic models overestimate the amount of anisotropy in the transverse flow \cite{Roma09}; this underlines the need for a relativistic theory of real fluids to really model relativistic heavy ion collisions.

\section{Real Fluids}

\subsection{Relativistic Real Fluids from Thermodynamics}
Let us return to our analysis of non relativistic fluids. Our starting point is eq. (\ref{dsdt}), where 
now we allow for entropy production, which must be positive. This suggests modifying the constitutive relations for an ideal fluid. We must take into account that $T^{ij}$must be symmetric, and the relativistic constraint $p^i=J_u^i/c^2$. This suggests writing

\be
T_{ij}=\frac12\left(p^jv^i+v^jp^i\right)+p\delta^{ij}+\tau_{ij}
\te
where

\be
\tau_{ij}=-\eta\sigma_{ij}-\xi\delta_{ij}\nabla_kv^k
\te 

\be 
\sigma_{ij}=\nabla_iv_j+\nabla_jv_i-\frac 23\delta_{ij}\nabla_kv^k
\te 
and

\be
J^ j_u=\left(\rho+p\right)v^j+\frac12\left(p^jv^2-v^j\left(v_ip^i\right)\right)+v_i\tau^{ij}
\te

\be 
\chi_j=-\kappa\nabla_j\alpha
\te 
In what follows we assume $\xi=0$. With these constitutive relations we find

\be 
\frac{\partial s}{\partial t}=-\nabla_jS^j+{\kappa}\left(\nabla_j\alpha\right)^2 +\frac{\eta}T\sigma^{ij}\sigma_{ij}+\frac{\xi}T\left(\nabla_jv^j\right)^2 +\frac 1T\left(\vec{p}\times\vec{v}\right)\nabla\times\vec{v}
\te
The last term vanishes in the rest frame and will not affect our argument.

To find the relativistic generalization of this theory, we observe that in the rest frame

\be
T^{\mu\nu}=\left(
\begin{array}{cc}
\rho&0\\
0&p\delta^{ij}+\tau^{ij}
\end{array}\right)
\te
Thus in any frame we may write

\be
T^{\mu\nu}= T^{\mu\nu}_0+\Pi^{\mu\nu}
\te
where $T^{\mu\nu}_0=\rho u^{\mu}u^{\nu}+ p\Delta^{\mu\nu}$ and
\be
\Pi^{\mu\nu}=-\eta\sigma^{\mu\nu}
\te
where $\sigma$ is the shear tensor

\be
\sigma_{\mu\nu}=\Delta^{\lambda}_{\mu}\Delta^{\rho}_{\nu}\left[u_{\rho,\lambda}+u_{\lambda,\rho}-\frac23\Delta_{\lambda\rho}u^{\alpha}_{,\alpha}\right]
\te
This is the so-called Chapman-Enskog prescription, and although it leads to a covariant generalization of the Navier-Stokes equations, it is untenable.

The conservation equations read
\be
T^{\mu\nu}_{0,\nu}+\Pi^{\mu\nu}_{,\nu}=0
\te
Now

\be
T^{\mu\nu}_{0,\nu}=\rho_{,\nu}u^{\mu}u^{\nu}+\left(\rho+p\right)u^{\mu}_{,\nu}u^{\nu}+\left(\rho+p\right)u^{\mu}u^{\nu}_{,\nu}+\Delta^{\mu\nu}p_{,\nu}=0
\te
Therefore

\bea
\dot{\rho}+\left(\rho+p\right)u^{\nu}_{,\nu}-u_{\mu}\Pi^{\mu\nu}_{,\nu}&=&0\nn
\dot{u}^{\mu}+\Delta^{\mu\nu}\frac{p_{,\nu}+\Pi^{\lambda}_{\nu,\lambda}}{\left(\rho+p\right)}&=&0
\tea
Moreover, since $\Pi^{\mu\nu}$ is both traceless and transverse, we may write the first equation as

\be
\dot{\rho}+\left(\rho+p\right)u^{\nu}_{,\nu}+\frac12\Pi^{\mu\nu}\sigma_{\mu\nu}=0
\te

\subsection{The causality problem}
Since in the Chapman-Enskog approach the viscous energy - momentum tensor (VEMT) is slaved to the four velocity, the only dynamic degrees of freedom are the components of the four velocity themselves. Let us consider the linear perturbations around a solution with $\beta^{\mu}=$ constant. Let us  write $u^ i=un^ i$, $n^ in_i=1$ and $u^0=\sqrt{u^ 2+1}$. We seek solutions that depend on a single variable $k_{\mu}x^{\mu}=k_ix^ i-\omega t$. We further decompose $k^i=Kn^i+k^ i_{\perp}$, $x^ i=xn^ i+x^i_{\perp}$. Since the theory is covariant, we may always perform a Lorentz transformation to the rest frame, namely we introduce new time and space variables

\bea 
t'&=&u^0t-ux\nn
x'&=&u^ 0x-ut
\tea
and new frequency and wave number

\bea 
\omega'&=&u^0\omega-uK\nn
K'&=&u^ 0K-u\omega
\tea
with $x'^i_{\perp}=x^i_{\perp}$ and $k'^i_{\perp}=k^i_{\perp}$. We seek solutions where $k^ i$ is real and $\mathrm{Im}\left[\omega\right]\le 0$.

In this frame $u'^ i=0$, $u'^ 0=1$ and $T'=T$. The linearized perturbations must have $\delta u'^0=0$. For the other variables we seek a solution 

\bea 
\delta' T'=T\delta' e^{-i\omega' t'+ik'_jx'^j} \nn
\delta' u'^i=\delta'^ i e^{-i\omega' t'+ik'_jx'^j}
\tea
The equations of motion become

\bea
-i\omega'\delta'+ic^2k'_j\delta'^{j}&=&0\nn
-i\omega'\delta'^{j}+ik'^j\delta'+{\gamma}\left(\left( K'^2+k_{\perp}^2\right)\delta'^{j}+\frac 13k'^jk'_k\delta'^{k}\right) &=&0
\tea
where $\gamma =\eta/\rho +p$. For a transverse perturbation $k'_k\delta'^{k}=0$, $\delta'=0$ we must have $\omega'=-i\gamma \left( K'^2+k_{\perp}^2\right)$. We observe that because $u^0>0$, the imaginary parts of $\omega$ and $\omega'$ have the same sign, so we can check stability by checking the imaginary part of $\omega'$.
Under a Galilean transformation $K'=K$ and $\omega'=\omega-uK$, so the theory is stable in all frames. However, under a relativistic transformation $K'=\left(  K-u\omega'\right) /u^0$. If $u\not= 0$ we get a quadratic equation for $\omega'$. For $k_{\perp}=0$ this is

\be 
\omega'^2-2\left(\frac Ku+\frac{iu^{02}}{2\gamma u^2} \right)\omega' +\frac{K^2}{u^2}=0
\te 
with solution

\be 
\omega'=\frac Ku\left[ 1+\frac{iu^{02}}{2\gamma Ku}\pm\sqrt{\left(1+\frac{iu^{02}}{2\gamma Ku} \right)^2-1 }\right] 
\te
When $u\to 0$, one solution is the expected one $\omega'=-i\gamma K^ 2$ but the second solution becomes $\omega'=iu^{02}/\gamma u^2$ and is unstable. For a longitudinal perturbation $\delta'^j=k'^j\delta''$ and the dispersion relation is

\be 
\omega'^2-c^2\left( K'^2+k_{\perp}^2\right)+\frac 43i\gamma \omega'\left( K'^2+k_{\perp}^2\right)=0
\te
The same analysis as before leads to a cubic equation. When $u\to 0$ two roots correspond to damped sound waves; the third root $\omega'\approx 3i/4\gamma u^2$ is unstable.

\subsection{Relativistic kinetic theory}
In the kinetic theory description \cite{Isr72} the transport equation reads 

\be
p^{\mu}\partial_{\mu}f=\frac{-1}{\tau}\mathrm{sign}\left(p^0\right)I_{col}
\te
$\tau$ is the so-called relaxation time. The current is

\be
J^{\mu}=e\int\;Dp\;p^{\mu}f
\te
and the EMT is

\be
T^{\mu\nu}=\int\;Dp\;p^{\mu}p^{\nu}f
\te
where 

\be
Dp =\frac{2d^4p\delta\left(p^2\right)}{\left(2\pi\right)^3}=\frac{d^4p}{\left(2\pi\right)^3 p}\left(\delta\left(p^0-p \right) + \delta\left(p^0+p \right)\right) 
\te
For simplicity we assume massless particles. We do not assume particle number conservation. To enforce energy-momentum conservation  we require

\be
\int\;Dp\;p^{\mu}\mathrm{sign}\left(p^0\right)I_{col}=0
\te
assuming for simplicity Maxwell - Boltzmann statistics, the equilibria are of the form

\be
f_0=\exp\left\{-\left|\beta_{\mu}p^{\mu}\right|\right\}
\te
For a given $T^{\mu\nu}$ we can always find a  local equilibrium distribution  $f_0$ such that the ideal fluid energy momentum tensor built from it

\be 
T_0^{\mu\nu}=\int\;Dp\;p^{\mu}p^{\nu}f_0=\rho u^{\mu}u^{\nu}+p\Delta^{\mu\nu}
\label{EMT0}
\te 
obeys

\be 
T_0^{\mu\nu}u_{\nu}=-\rho u^{\mu}
\label{LL0}
\te
with the same energy density and four velocity as from the Landau-Lifshitz prescription. We then define a temperature from

\be
\rho=\sigma T^4
\te
where $\sigma=\pi^2/15\approx 0.66$. Using the Maxwell - Juttner distribution rather than the Bose-Einstein one is equivalent to the approximation

\be 
\sum_{n=1}n^{-4}=\frac{\pi^4}{90}\approx 1
\te
whereby $\sigma$ becomes $6/\pi^2\approx 0.61$. We make this approximation from now on.

It follows that the viscous energy momentum tensor

\be 
\Pi^{\mu\nu}=T^{\mu\nu}-T_0^{\mu\nu}
\label{VEMT}
\te 
is traceless and transverse

\be 
\Pi^{\mu\nu}u_{\nu}=0
\label{TTC}
\te 
We parametrize

\be 
f=f_0\left[1+Z\right]
\te 
with $Z=0$ at equilibrium. The transversality condition becomes

\be
\int\;Dp\;p^{\mu}\left(-u_{\nu}p^{\nu}\right)f_0Z=0
\label{trans}
\te
We assume a simple Boltzmann type entropy flux \cite{Sag13}

\be
S^{\mu}=-\int\;Dp\;\left(\mathrm{sign}\left(p^0 \right)\right)p^{\mu}f\left[\ln \frac{f}{f_0}-1\right]
\te
we get the entropy production

\be
S^{\mu}_{,\mu}=\frac{1}{\tau}\int\;Dp\;I_{col}\ln\left[1+Z\right] 
\te

\subsection{Chapman-Enskog and Grad}

The Chapman-Enskog procedure seeks a formal expansion for $Z$ in powers of $\tau$. To this end, one parametrizes

\be
Z=\tau Z_1+\tau^2Z_2+\ldots
\te
The ``spatial'' derivatives $\Delta^{\mu\nu}T_{,\nu}$ and $\Delta^{\mu\nu}u^{\lambda}_{,\nu}$ are regarded as zeroth order quantities, while the ``time'' derivatives $\dot{T}=u^{\mu}T_{,\mu}$ and
$\dot{u}^{\lambda}=u^{\mu}u^{\lambda}_{,\mu}$ are derived from energy-momentum conservation, which is also a necessary consistency condition. 

To find $Z_1$ we only need the linearized collision integral. This must be a symmetric operator, it must obey the energy momentum conservation constraint, must lead to non negative entropy production and must admit thermal distributions as the only homogeneous distributions. To satisfy these requirements, we write 

\be 
I_{col}\left( p\right) =F\left[p\right]  f_0\left( p\right) \left[Z\left(p \right) -\int\;Dp'\;K\left[ p,p'\right] F\left[p'\right]  f_0\left( p'\right) Z\left( p'\right) \right] 
\te
The second term is there to enforce the constraints, namely

\be
\int\;Dp\;\mathrm{sign}\left(p^0\right)p^{\mu}F  f_0Z=\int\;DpDp'\;\mathrm{sign}\left(p^0\right)p^{\mu}Ff_0K\left[ p,p'\right] F' f'_0Z' 
\te
Since the kernel $K$ is symmetric and these hold for any $Z$ we must have

\be
\int\;Dp'\;K\left[ p,p'\right] \mathrm{sign}\left(p'^0\right)p'^{\mu}F' f'_0=p^{\mu}\mathrm{sign}\left(p^0\right)
\te
It also makes $I_{col}$ to vanish when $Z$ is just a variation in $\beta_{\mu}$.

We adopt the Anderson - Witting prescription $F=\left|-u_{\mu}p^{\mu}\right|$ \cite{AndWit74,TakInu10}. Write

\be 
K \left[ p,p'\right] =K_{\rho\sigma}p^{\rho}\mathrm{sign}\left(p^0\right)p'^{\sigma}\mathrm{sign}\left(p'^0\right)
\te
Then

\be
 K_{\rho\sigma}I^{\sigma\mu}=\delta^{\mu}_{\rho}
\te
where

\be
I^{\sigma\mu}=u_{\lambda}\int\;Dp\;\mathrm{sign}\left(p^0\right)p^{\sigma}p^{\mu}p^{\lambda} f_0=A_3\left[u^{\sigma}u^{\mu}+\frac13\Delta^{\sigma\mu}\right]
\te

\be
A_3=\int\;Dp\;\left|-u_{\mu}p^{\mu}\right|^3 f_0=\frac{24}{\pi^2}T^5
\te
This means that

\be
K_{\rho\sigma}=A_3^{-1}\left[u_{\rho}u_{\sigma}+3\Delta_{\rho\sigma}\right]
\te
Finally the collision integral is

\be 
I_{col}\left( p\right) =\left| -u_{\mu}p^{\mu}\right|   f_0\left( p\right) \left[Z\left(p \right) -K_{\rho\sigma}p^{\rho}\mathrm{sign}\left(p^0\right)\int\;Dp'\;p'^{\sigma}\left(-u_{\mu}p'^{\mu}\right)  f_0\left( p'\right) Z\left( p'\right) \right] 
\te
If $Z$ satisfies the constraint eq. (\ref{trans}), then the second term is zero and the entropy production is positive, provided $Z\ge -1$.

To first order

\be
Z_1=-\frac{p^{\mu}p^{\nu}}{\left|-p^{\alpha}u_{\alpha}\right|}\beta_{\nu,\mu}
\label{foc}
\te
Now in general

\bea
\beta_{\nu,\mu}&=&\frac1T\left[u_{\nu,\mu}-u_{\nu}\frac{T_{,\mu}}{T}\right]\nn
&=&\frac1T\left[u_{\mu}u_{\nu}\frac{\dot{T}}{T}+\frac12\left[\sigma_{\mu\nu}+\frac23\Delta_{\mu\nu}u^{\lambda}_{,\lambda}\right]-\frac12\left[u_{\nu}\Delta^{\lambda}_{\mu}+u_{\mu}\Delta^{\lambda}_{\nu}\right]\left(\frac{T_{,\lambda}}{T}+\dot{u}_{,\lambda}\right)\right.\nn
&-&\left.\frac12\left[u_{\nu}\Delta^{\lambda}_{\mu}-u_{\mu}\Delta^{\lambda}_{\nu}\right]\left(\frac{T_{,\lambda}}{T}-\dot{u}_{,\lambda}\right)+\frac12\Delta^{\lambda}_{\nu}\Delta^{\rho}_{\mu}\left[u_{\lambda,\rho}-u_{\rho,\lambda}\right]\right]
\tea
so, using the zeroth order time derivatives and $c^2=1/3$

\be
Z_1=\frac{-1}{2T\left|p^{\alpha}u_{\alpha}\right|}\sigma_{\mu\nu}{p^{\mu}p^{\nu}}
\label{trans2}
\te
The viscous energy momentum tensor is

\be 
\Pi_1^{\mu\nu}=\frac{-\tau}{2T}\int\;Dp\;f_0\frac{p^{\mu}p^{\nu}p^{\rho}p^{\lambda}}{\left|p^{\alpha}u_{\alpha}\right|}\sigma_{\rho\lambda}=-\eta\sigma^{\mu\nu}
\label{VEMTCE1}
\te 
where

\be
\eta=\frac{\tau}{15T}\int\;Dp\;f_0\left|-p^{\alpha}u_{\alpha}\right|^3=\frac{8}{5\pi^2}\tau T^4
\te
The Chapman-Enskog procedure cannot generate a true dynamical equation for the viscous EMT, and thus does not solve the causality/stability problems. The Grad approach takes the different strategy of keeping the form eq. (\ref{trans2}), but replacing $\sigma_{\mu\nu}$ by a new tensor $C_{\mu\nu}$ regarded as a new independent variable.

\be
Z_G=\frac{-1}{2T\left|-p^{\alpha}u_{\alpha}\right|}C_{\mu\nu}{p^{\mu}p^{\nu}}
\label{Gradansatz}
\te
 $C_{\mu\nu}$ is defined up to  a multiple of $g_{\mu\nu}$, so we may assume it is traceless, and because of the constraint eq. (\ref{trans}) it must be transverse. 
The viscous energy momentum tensor becomes

\be 
\Pi_G^{\mu\nu}=\frac{-1}{2T}\int\;Dp\;f_0\frac{p^{\mu}p^{\nu}p^{\rho}p^{\lambda}}{\left|-p^{\alpha}u_{\alpha}\right|}C_{\rho\lambda}=-\frac{\eta}{\tau}C^{\mu\nu}
\label{VEMTG}
\te
The next step is to substitute this into the Boltzmann equation. The nonlocal term vanishes and we get 

\be
\frac{1}{\tau}\left(p^{\nu}u_{\nu}\right)f_0Z=\frac{\partial}{\partial x^{\mu}}\left[p^{\mu}f_0\left(1+Z \right) \right]
\label{BE2}
\te
We wish to extract from there an equation for $C^{\mu\nu}$. The time-honored procedure is to consider the moments of this equation. Since $Z$ is even the zeroth moment vanishes, and the first moment gives back energy - momentum conservation.

\be
\frac{\partial}{\partial x^{\lambda}}\int\;Dp\;p^{\rho}p^{\lambda}f_0\left(1+Z \right) =0
\te
Therefore the first non trivial choice is to use the second moments

\be
\frac{\partial}{\partial x^{\mu}}\int\;Dp\;\mathrm{sign}\left(p^0\right)p^{\rho}p^{\lambda}p^{\mu}f_0\left(1+Z \right) =\frac{-1}{\tau}\int\;Dp\;p^{\rho}p^{\lambda}\left|-p^{\nu}u_{\nu}\right|f_0Z
\te
To evaluate these equations, we use the following identities

\bea 
\int\;Dp\;p^{\rho}p^{\lambda}f_0&=&{A_2}\left[u^{\rho}u^{\lambda}+\frac 1{3}\Delta^{\rho\lambda}\right]\nn
\int\;Dp\;\mathrm{sign}\left(p^0\right)p^{\rho}p^{\lambda}p^{\mu}f_0&=&{A_3}\left[u^{\rho}u^{\lambda}u^{\mu}
+\frac 1{3}\left( \Delta^{\rho\lambda}u^{\mu}+\Delta^{\mu\lambda}u^{\rho}+\Delta^{\mu\rho}u^{\lambda}\right) \right]\nn
\int\;Dp\;p^{\rho}p^{\lambda}p^{\mu}p^{\nu}f_0&=&{A_4}\left\lbrace u^{\rho}u^{\lambda}u^{\mu}u^{\nu}+\frac 1{3}\left( \Delta^{\rho\lambda}u^{\mu}u^{\nu}+...\right) +\frac 1{15}\left( \Delta^{\rho\lambda}\Delta^{\mu\nu}+...\right)\right\rbrace\nn
\int\;Dp\;\mathrm{sign}\left(p^0\right)p^{\rho}p^{\lambda}p^{\mu}p^{\nu}p^{\theta}f_0&=&  {A_5}\left\lbrace u^{\rho}u^{\lambda}u^{\mu}u^{\nu}u^{\theta}+\frac 1{3}\left( \Delta^{\rho\lambda}u^{\mu}u^{\nu}u^{\theta}+...\right) +\frac 1{15}\left( \Delta^{\rho\lambda}\Delta^{\mu\nu}u^{\theta}+...\right)\right\rbrace \nn
\int\;Dk\;k^{\rho}k^{\lambda}k^{\mu}k^{\nu}k^{\theta}k^{\phi}F\left[\left(k^{\alpha}u_{\alpha}\right) \right]&=&  {A_6}\left\lbrace u^{\rho}u^{\lambda}u^{\mu}u^{\nu}u^{\theta}u^{\phi}+\frac13\left( \Delta^{\rho\lambda}u^{\mu}u^{\nu}u^{\theta}u^{\phi}+...\right)\right. \nn
 &+&\left. \frac1{15}\left( \Delta^{\rho\lambda}\Delta^{\mu\nu}u^{\theta}u^{\phi}+...\right)+\frac1{105}\left( \Delta^{\rho\lambda}\Delta^{\mu\nu}\Delta^{\theta\phi}+...\right)\right\rbrace 
\tea 
In the following, we shall adopt
\be 
A_a=\int\;Dp\;\left|-p^{\alpha}u_{\alpha}\right|^a f_0=\left(a+1 \right) !\frac{T^{a+2}}{ \pi^2 }
\te 
The equation becomes

\be 
\frac{\partial}{\partial x^{\mu}}{T^5}C^{\rho\lambda\mu}=\frac{T^5}{\tau}C^{\rho\lambda}
\te
where

\be 
C^{\rho\lambda\mu}= \left\{3u^{\rho}u^{\lambda}u^{\mu}
+ \Delta^{\rho\lambda}u^{\mu}+\Delta^{\mu\lambda}u^{\rho}+\Delta^{\mu\rho}u^{\lambda}-u^{\lambda}C^{\rho\mu}- u^{\rho}C^{\mu\lambda}-u^{\mu}C^{\rho\lambda} \right\}
\te
It is not really possible to kill all second moments simultaneously. We shall be happy to kill the transverse traceless contribution, namely

\be 
\left[\Delta_{\tau\rho}\Delta_{\sigma\lambda}-\frac13\Delta_{\tau\sigma}\Delta_{\rho\lambda}\right]\frac{\partial}{\partial x^{\mu}}T^{5}C^{\rho\lambda\mu}=\frac{T^5}{\tau}C_{\tau\sigma}
\te 
which reduces to

\bea 
\frac {-1}{T^5}\frac{\partial}{\partial x^{\mu}}{T^{5}}u^{\mu}C_{\tau\sigma}&-&\Delta_{\sigma\lambda}C^{\rho\lambda\mu}\frac{\partial u_{\tau}u_{\rho}}{\partial x^{\mu}}
-\Delta_{\tau\rho}C^{\rho\lambda\mu}\frac{\partial u_{\sigma}u_{\lambda}}{\partial x^{\mu}}\nn
&+&\frac13\Delta_{\tau\sigma}C^{\rho\lambda\mu}\frac{\partial u_{\rho}u_{\lambda}}{\partial x^{\mu}}+\frac13\Delta_{\rho\lambda}C^{\rho\lambda\mu}\frac{\partial u_{\sigma}u_{\tau}}{\partial x^{\mu}}=\frac{1}{\tau}C_{\tau\sigma}
\label{GE}
\tea 
Keeping only linear terms, this is an equation of Maxwell - Cattaneo type for $C_{\mu\nu}$ \cite{Max67,JosPre89}

\be
\dot{C}_{\tau\sigma}+\frac{1}{\tau}C_{\tau\sigma}-\sigma_{\tau\sigma}+\mathrm{ho}=0
\te 
Actually, we may get an expression for the time derivative of $Z$ directly from the transport equation and then use it to find an equation for the viscous energy momentum tensor \cite{DMNR12}. 
However, beyond the leading order this equation involves integrals which cannot be simple expressed in terms of the known moments of the distribution function. Therefore to obtain a definite equation we must provide a \emph{closure}, i. e., an expression for $Z$ allowing us to compute the integrals. For example, if we use the Grad ansatz $Z_G$ we get an equation with the same structure as eq. (\ref{GE}), though the coefficients may be different.

We wish to check that the Grad approach is consistent with causality and stability \cite{HisLin83,Ols90,OlsHis90}. Introducing a new perturbation  for $C^{ij}$ we get the equations

\bea
-i\omega'\delta'+ic^2k'_j\delta'^j&=&0\nn
-i\omega'\delta'^i+ik'^i\delta' -i\frac{\gamma}{\tau}k'_jC'^{ij}&=&0\nn
-i\omega'C'^{ij}+\frac{1}{\tau}C'^{ij}-i\left( k'^i\delta'^j+k'^j\delta'^i-\frac 23\delta^{ij}k'_k\delta'^k\right) &=&0
\tea
We revert to the Chapman - Enskog equations with the replacement $\gamma\to\gamma /1-i\omega'\tau$. For transverse perturbations we now get a quadratic equation

\be 
\left(1-i\omega'\tau\right)\omega'+i\gamma K'^2=0
\te 
It is easy to see that both roots are stable when $u=0$. For general $u$ stability obtains if $\tau\ge\gamma$. It is interesting to observe that with the expressions above $\tau\approx 5\gamma$.

We now consider longitudinal waves with $k_{\perp}=0$. The dispersion relation is

\be 
\left(1-i\omega'\tau\right)\left[\omega'^2-\frac{c^2}{u^2_0}\left(  K-u\omega'\right)^2\right]+\frac 4{3u^2_0}i\gamma \omega'\left(  K-u\omega'\right)^2=0
\te
We rearrange this as

\be
\omega'\left[\omega'^2-C^2\left(  K-u\omega'\right)^2\right]
+\frac{i}{\tau}\left[\omega'^2-\frac{c^2}{u^2_0}\left(  K-u\omega'\right)^2\right]=0
\te
where 

\be
C^2=\frac{c^2}{u^2_0}\left(1+\frac {4\gamma}{3c^2\tau}\right)
\te
This is

\be
\omega'\left(\omega'-\frac{CK}{\left(1+Cu\right)}\right)\left(\omega'+\frac{CK}{\left(1-Cu\right)}\right)
+\frac{i}{\left(1-C^2u^2\right)\tau}\left[\omega'^2-\frac{c^2}{u^2_0}\left(  K-u\omega'\right)^2\right]=0
\te
In the formal limit $\tau\to\infty$ we find three real roots $\omega'=0,\pm\omega_{\pm}$, where $\omega_{\pm}= CK/1\pm Cu$. At large but finite $\tau$, the solution that goes to zero behaves as

\be
\omega_0=\frac{-i}{\tau}\frac{c^2}{u^2_0C^2}
\te
and it is stable. The solutions which converge to $\pm\omega_{\pm}$ behave as

\be
2CK\omega_{\pm}\left(\omega'-\omega_{\pm}\right)+\frac {4i\gamma}{3u^2_0\tau^2}\left(  K\mp u\omega_{\pm}\right)^2=0
\te
so they are stable too. Of course, we also have solutions with $\delta=\delta^i=k'_jC^{ij}=0$ and $\omega'=-i/\tau$, which are obviously stable. 

\subsection{Entropy production variational method}

The Grad approach as we have presented it still has the problems that the Grad ansatz eq. (\ref{Gradansatz}) does not lead to a non negative one particle distribution function (quite the opposite, since $C_{\mu\nu}{p^{\mu}p^{\nu}}$ must be negative in some direction in momentum space) and that it is unclear how to introduce nonlinear terms. To overcome these problems we need a better motivated closure for $Z$. We shall try to find it by seeking the value of $Z$ which minimizes entropy production for a given VEMT, and satisfies the constraint eq. (\ref{TTC}) \cite{CalPR10}. Adding Lagrange multipliers $\zeta_{\mu\nu}$  we get the equation

\be
\left\{I_{col}\left[\ln\left[1+Z\right]\right] +\frac{I_{col}\left[Z\right] }{1+Z}\right\}= 
-\tau\zeta_{\mu\nu} p^{\mu}p^{\nu}f_0
\label{EPVM}
\te
The point is that this equation has bounded solutions for any value of the right hand side. We may also regard it as a means to obtain a formal series solution for $Z$ in powers of $\tau$, where, if we keep only the first order term, we recover the Grad ansatz. 

It is convenient to introduce a new unknown $\chi=\ln\left[1+Z\right]$, with inverse transformation $Z=e^{\chi}-1$. In terms of the new unknown, the equation reads 

\be
I_{col}\left[\chi\right]= 
-\frac{\tau}2\zeta_{\mu\nu} p^{\mu}p^{\nu}f_0-\frac12\left[e^{-\chi}I_{col}\left[e^{\chi}-1\right]-I_{col}\left[\chi\right]\right]
\label{EPVM3}
\te
To lowest order we may neglect the second term in the right hand side. Observe that in any case $\zeta_{\mu\nu}$ is defined up to a multiple of $g_{\mu\nu}$. We use this freedom to require $\zeta_{\mu\nu}u^{\mu}u^{\nu}=0$. Transversality requires

\be
0=\int\;Dp\;\mathrm{sign}\left(p^0\right)p^{\lambda}f_0\zeta_{\mu\nu} p^{\mu}p^{\nu}=\zeta_{\mu\nu}A_3\left[u^{\lambda}u^{\mu}u^{\nu}+\frac13\left(u^{\lambda}\Delta^{\mu\nu}+u^{\mu}\Delta^{\lambda\nu}+u^{\nu}\Delta^{\lambda\mu}\right)\right]
\te
Therefore $\zeta_{\mu\nu}$ must be traceless and transverse.  To lowest order 

\be
\chi=\frac{-1}2\frac{\zeta^{\left(0\right)}_{\mu\nu} p^{\mu}p^{\nu}}{\left| -u_{\rho}p^{\rho}\right|}
\te 
Let us further expand the exponential. In the rest frame  the energy momentum tensor reads

\bea
T^{00}&=&\sigma T^4\equiv \rho\nn
T^{0i}&=&0\nn
T^{ij}&=&\frac13 \sigma T^4 \delta^{ij}-\frac{A_3}{15}\tau\zeta^{ij}
\label{tfromz}
\tea
The kinetic equation to lowest order in $\tau$ reads

\be
p^{\lambda}\partial_{\lambda}f_0\left[1-\tau\frac{1}2\frac{\zeta_{\mu\nu} p^{\mu}p^{\nu}}{\left| -u_{\rho}p^{\rho}\right|}\right]=\frac{1}2\left(  -u_{\mu}p^{\mu}\right)   f_0 \frac{\zeta_{\mu\nu} p^{\mu}p^{\nu}}{\left| -u_{\rho}p^{\rho}\right|}  
\te
The meaning of this equation is as a generating equation for its moments. The first order moments give energy momentum conservation. To find an equation for $\zeta_{\mu\nu}$ we go to the rest frame and multiply both sides by $\mathrm{sign}\left(p^0\right)p^ip^j$. Integrating and discarding the trace part, we get

\be 
\frac{A_4}{15 T}\sigma_{ij}-\frac{\tau A_5 \dot{T}}{15 T^2}\zeta_{ij}-\frac{\tau A_5}{105 T}\left[ u_{k,k}\zeta_{ij}+\zeta_{ik}\sigma ^k_j+\sigma_{i}^k\zeta_{kj}-\frac 23\Delta_{ij}\zeta^{\left(0\right)kl}\sigma_{kl}\right] =\frac{A_4}{15 } \left[\zeta_{ij}+\tau\zeta_{ij,0}\right] 
\te

\subsection{Non linear corrections to boost invariant flow}
We now turn to consider nonlinear corrections to this equations.

The strategy we are following consists on improving the stability of the theory by adding new variables obeying dynamical equations of their own. In many approaches, such as the so-called  Israel - Stewart theory \cite{is1,is2} or Extended Thermodynamics \cite{extended} the extra variables are the components of the viscous energy-momentum tensor $\Pi_{\mu\nu}$ itself. The dynamical equations for $\Pi_{\mu\nu}$ have been derived in a number of ways, such as carefully taking moments of the kinetic equation \cite{DMNR12}, a systematic gradient expansion of the kinetic theory \cite{BRW06}, from AdS-CFT correspondence \cite{cft} or simply writing down all terms consistent with the symmetries of the theory up to a certain order \cite{BRSSS}. We shall call these theories ``second order fluid dynamics'' (SOFD) for short, and take the presentation in \cite{EXG10} as a suitable representative.

In our approach (EPVM, or entropy production variational method), on the other hand, the new variables are the Lagrange multipliers enforcing the constraints on entropy production; $\Pi_{\mu\nu}$ itself may depend nonlinearly on the Lagrange multipliers. Our approach is therefore closer to the Geroch - Linblom Divergence type theories \cite{dtt}, although we shall not demand that the resulting theory conforms to the dissipative type framework.

To obtain further insight on the meaning of these theories, we shall apply them to the case of boost invariant flow. Adopting Milne coordinates $T^{\mu}_{\nu}$ is diagonal; we write $T^{\tau}_{\tau}=-\rho$, $T^{\eta}_{\eta}= \rho /3-\Pi$, $T^x_x=T^y_y=\rho/3+\Pi/2$. Energy-momentum conservation yields

\be 
\dot{\rho}+\frac1{\tau}\left( \frac 43\rho-\Pi\right) =0
\te
We also write $\zeta^{\left(0\right)\eta}_{\eta}=\zeta$, $\zeta^{\left(0\right)x}_{x}=\zeta^{\left(0\right)y}_{y}=-\zeta/2$. Recall that $\sigma^{\eta}_{\eta}=4/3\tau$, $\sigma^{x}_{x}=\sigma^{y}_{y}=-2/3\tau$. We get, discarding a derivative of the temperature,

\be 
\tau_R\dot{\zeta}+\zeta =\frac 4{3T\tau}-a_1 \tau_R\frac{\zeta}{\tau} 
\label{foeq}
\te 
where we have written $\tau_R$ for the relaxation time to avoid confusion with Milne time, and 

\be
a_1=\frac{ A_5}{3A_4T}
\te
The linearized collision term we are using is too simplistic to allow for a reliable derivation of nonlinear terms; however, in this case we may exploit the symmetries of the problem, which indicate that there is a single true degree of freedom $\zeta$ underlying the viscous energy momentum tensor. Therefore we generalize eq. (\ref{tfromz}) to 

\be 
\Pi=\frac{\tau_RA_3}{15}\zeta + h_1 \left( \frac{\tau_RA_3}{15}\zeta\right) ^2
\te 
and eq. (\ref{foeq}) to 

\be 
\tau_R\dot{\zeta}+\zeta =\frac 4{3T\tau}-a_1\tau_R \frac{\zeta}{\tau} - h_2\frac{\tau_RA_3}{15}\zeta^ 2
\label{soeq}
\te  

We determine the new transport coefficients $h_{1,2}$ by asking that, in a stationary situation, the relation between $\Pi$ and the shear $4/3\tau$, to second order in $\tau_R$, agrees with a systematic expansion for a Boltzmann gas \cite{EXG10}. Indeed, to second order we may write

\be 
\zeta = \frac{15}{\tau_RA_3}\left[ \Pi - h_1\Pi^2\right] 
\te 
and so, neglecting time derivatives

\be 
\Pi =\frac{\tau_R A_3}{15T}\frac 4{3\tau}-a_1\tau_R \frac{\Pi}{\tau}- \left( h_2-h_1\right) \Pi^2
\te 
Matching against SOFD yields \cite{EXG10}

\bea 
\tau_R &=&\frac{6\eta}{sT}\nn
\frac{A_3}{15T}&=&\frac{sT}{6}\nn
a_1&=&\frac 43\nn
h_2-h_1&=&\frac1{4\pi T\eta}
\tea
where $\eta$ is the shear viscosity and $s$ the entropy density, both extracted from a realistic equation of state. An analysis of the possible origins of the $h_{1,2}$ coefficients shows that both are of the same order of magnitude, $\approx\tau_R^{-1}$. When this holds, the behavior of the model is relatively insensitive to the actual values of $h_1$ and $h_2$.

\begin{center}

\begin{figure}[htb]

\scalebox{0.47}{\includegraphics{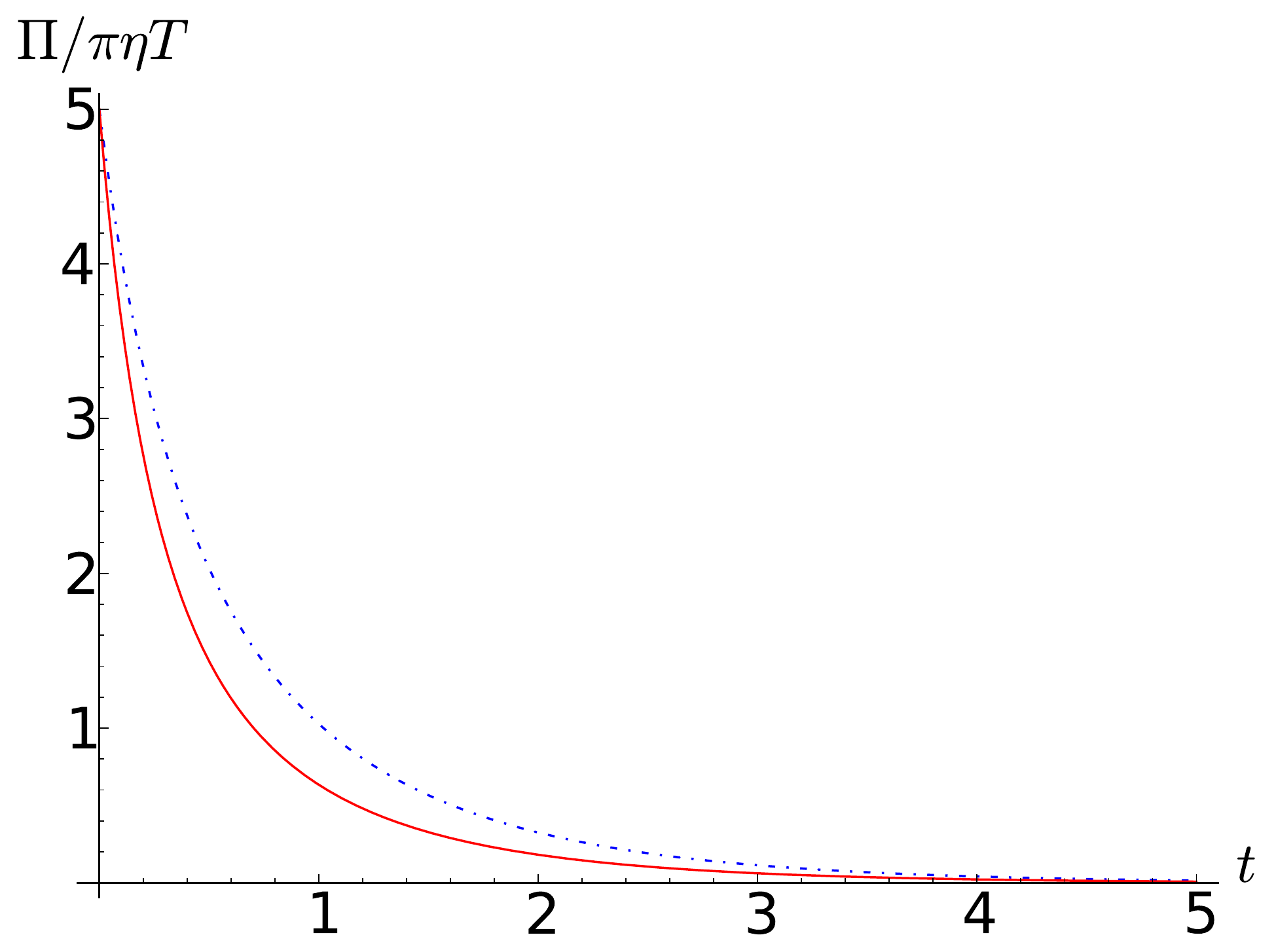}}

\caption{(Color online) The evolution of the viscous energy momentum tensor, as predicted by SOFD (dashes and dots) and EPVM (full line) starting form $x_0=5$. We see that EPVM predicts a faster approach towards $\Pi=0$. }

\label{decay}

\end{figure}

 \end{center}

To break the degeneracy between $h_1$ and $h_2$, let us consider a situation where the shear $\tau^{-1}$ terms are negligible and the transport coefficients are constant. The SOFD equation admits a stationary solution with $\Pi=\Pi_{0SOFD}=-1/\left( h_2-h_1\right) $; $\Pi$ is otherwise unbounded. The EPVM equations on the other hand imply a lower bound $\Pi\ge -1/\left( 4h_1\right) $, which is realized when $\zeta=-1/\left( 2h_1\eta T\right) $. There is a steady solution when $\zeta=-1/\left( h_2\eta T\right) $, which implies $\Pi=\Pi_{0EPVM}=(-1/h_2)\left[1-\left( h_1/h_2\right) \right]$. In both cases, the steady solutions are unstable. For SOFD, this implies the existence of runaway solutions, namely, solutions which begin below the fixed point (if it is negative) run away to minus infinity. In the EPVM, on the other hand, we may eliminate the runaway solutions by demanding that the steady solution coincides with the lower bound, and adopting, for each allowed value of $\Pi$, the value of $\zeta$ above the fixed point. This requires $h_1=h_2/2$, 
and therefore $h_2=2h_1=1/\left( 2\pi T\eta\right) $. Introducing the transport coefficient 

\be 
\lambda =\frac{\eta}{2\pi T}
\te 
The SOFD equation reads \cite{EXG10}

\be 
\tau_R\dot{\Pi}+\Pi =\frac {4\eta}{3\tau}-\frac 43\tau_R \frac{\Pi}{\tau}- \frac{\lambda}{2\eta^2} \Pi^2
\label{sofd}
\te 
while the extended EPVM yields the system \cite{CalPR10}

\bea 
\tau_R\dot{\zeta}+\zeta &=&\frac 4{3T\tau}-\frac 43\tau_R \frac{\zeta}{\tau} - \frac{\zeta^ 2}{2\pi}\nn
\Pi &=&\eta T\left[ \zeta + \frac{\zeta^2}{4\pi}\right] 
\tea
One way to visualize the difference between these models is to consider the free decay of $\Pi$. Suppose $\Pi$ is observed to have the value $\Pi_0$ at some time $\tau_0$ late enough that the $\tau^ {-1}$ terms in the equations may be neglected, and the transport coefficients regarded as constant. Then, according to SOFD, the further decay of the VEMT is given by

\be 
\frac{\Pi}{\Pi_0}=\frac{1}{\left( 1+\frac{x_0}{4}\right)e^{t}-\frac{x_0}{4} }
\te 
where $t=\tau-\tau_0/\tau_R$ and $x=\Pi/\pi\eta T$, while from EPVM we get

\be 
\zeta=\frac{\zeta_0}{\left( 1+\frac{\zeta_0}{2\pi}\right)e^{t}-\frac{\zeta_0}{2\pi} }
\te 
where

\be 
\frac{\zeta_0}{2\pi}=\sqrt{1+x_0}-1
\te
Besides the absence of runaway solutions already noted, the EPVM provides for a faster decay of large fluctuations (see fig. (\ref{decay})). 

\section{Final remarks}

The theory of relativistic real fluids has a curious history because while the stability problems we have discussed have been known for a long time \cite{JosPre89} , yet they do not seem to have elicited any strong response until fairly recently. There were known ways to improve the theory (foremost the Israel - Stewart \cite{is1,is2} and extended thermodynamics theories \cite{extended}), and also a family of theories which were known to be free of such problems on a rigorous basis (the Geroch-Lindblom dissipative type theories \cite{dtt}). However, the former were presented as successive approximations to a yet unknown theory, and the physical foundations of the latter remained elusive \cite{dtt2}. It was only the realization that relativistic real fluids might be produced in RHICs that triggered an all-out attack on the problem, to the extent that it would be impossible to describe all this activity in a short review such as this. We have therefore aimed to present just the fundamental ideas behind the theory, what the main problems are, and which lines of thought seem to us likely to be fruitful, and this almost entirely from the formal side, leaving phenomenology to more knowledgeable authors.

Concerning the present state of the theory, it seems fair to say that we have a reliable understanding of evolution during the hydrodynamic era, and the beginnings of a theory of the freeze out transition. The early times of the collision are relatively much more poorly understood. In particular, we do not know how hydrodynamic behavior may arise on such short time scales as demanded by theory, though the work on non-abelian instabilities \cite{inst} and QGP turbulence \cite{turb} , on one hand, and on AdS-CFT correspondence on the other \cite{cfttherm}, makes those scales look not so unrealistic as they used to. 

We can only conclude that we are only witnessing  the early childhood of the relativistic real fluids - RHICs connection, and this is what makes this such an exciting field to work on. 

\section*{Acknowledgment}
This work has been developed in collaboration with Jer\'onimo Peralta Ramos. It is supported in part by Universidad de Buenos Aires, CONICET and ANPCYT (Argentina)

\section*{ Appendix: Kadanoff-Baym equations and quantum kinetic field theory}

In this appendix we shall discuss the derivation of kinetic theory from quantum field theory. The presentation follows \cite{CalHu08}.

The CTP generating functional depends on two external sources

\begin{equation}
e^{iW\left[J^{1},J^{2}\right]}=\int D\Phi^{1}D\Phi^{2}\:e^{i\left\{S\left[\Phi^{1}\right]-S\left[\Phi^{2}\right]+\int \left(J^{1}\Phi^{1}-J^{2}\Phi^{2}\right)\right\}}
\label{five}
\end{equation}

It defines two background fields through

\begin{equation}
\phi^{1}\left[x\right]=\frac{\delta W\left[J^{1},J^{2}\right]}{\delta J^{1}\left[x\right]};\ \ \phi^{2}\left[x\right]=-\frac{\delta W\left[J^{1},J^{2}\right]}{\delta J^{2}\left[x\right]}
\label{six}
\end{equation}



The CTPEA is the full Legendre transform 

\begin{equation}
\Gamma\left[\phi^1,\phi^2\right]=W\left[J^{1},J^{2}\right]-\int \left(J^{1}\phi^{1}-J^{2}\phi^{2}\right)
\label{seven}
\end{equation}

It generates the equations of motion 

\begin{equation}
\frac{\delta \Gamma\left[\phi^1,\phi^2\right]}{\delta\phi^1\left[x\right]}=-J^1\left[x\right];\ \ \frac{\delta \Gamma\left[\phi^1,\phi^2\right]}{\delta \phi^2\left[x\right]}=J^2\left[x\right]
\label{eight}
\end{equation}

The equation of motion for the mean field is obtained when $J^1=J^2$ by setting $\phi^1=\phi^2$ in the equations (\ref{eight}) \emph{after} computing the variational derivatives.



To compute the CTPEA, observe that

\begin{equation}
\Gamma\left[\phi^1,\phi^2\right]=S\left[\phi^1\right]-S\left[\phi^2\right]+\mathrm{quantum\: corrections}
\label{nine}
\end{equation}

The quantum corrections are the sum of all the one-particle irreducible (1PI) graphs in the theory.
The linearized one-particle irreducible ($%
1PI$) effective action has the structure

\begin{eqnarray}
\Gamma _{1PI} &=&\int d^{d}xd^{d}y\;\left\{ \varphi_- \left( x\right) \left[D\left(
x,y\right) +\mathbf{D}\left( x,y\right) \right] \varphi_+\left( y\right) \right.  \nonumber \\
&&\left. +\frac{i}{2}\varphi_- \left( x\right) \mathbf{N}%
\left( x,y\right) \varphi_-\left( y\right)
\right\}
\end{eqnarray}
$\varphi_-=\left[ \varphi
^{1}-\varphi ^{2}\right]$, $\varphi_+=\left[ \varphi
^{1}+\varphi ^{2}\right]/2$.

\begin{equation}
D(x,y)=\left[ \partial _{x}^{2}-m_{b}^{2}\right] \delta (x-y)
\end{equation}
$\mathbf{D}$ is causal and $\mathbf{N}$ is even, and both are real. 
A good deal of our discussion will revolve around the different properties
of the propagators of the theory, that is, the expectation values of binary
products of field operators with respect to the initial state. Since field
operators at different locations do not generally commute, we have several
different propagators according to the ordering of the field operators
within the expectation value. 

The equations of motion for the propagators are derived from the
identity

\begin{equation}
\frac{{\cal D} ^{2}\Gamma _{1PI}}{{\cal D} \varphi ^{a}{\cal D} \varphi ^{b}}G^{bc}%
=i\hbar \delta^c_{a}
\end{equation}


The $G^{ab}$ above denote the four basic propagators
\medskip

Feynman $G_{F}\equiv <T\left( \Phi \left( x\right) \Phi \left( x^{\prime
}\right) \right) >=G^{11}$,
\medskip

Dyson $G_{D}\equiv <\tilde{T}\left( \Phi \left(
x\right) \Phi \left( x^{\prime }\right) \right) >=G^{22}$,
\medskip

Positive frequency $G^{+}\equiv <\Phi \left( x\right) \Phi \left( x^{\prime }\right) >=G^{21}$,
\medskip

Negative frequency $G^{-}\equiv <\Phi \left( x^{\prime }\right) \Phi \left(
x\right) >=G^{12}$,
\medskip

where $T$ stands for time ordering and $\tilde{T}$ stands for
anti-time ordering.

Explicitly

\bea
\left[D+\mathbf{D}_{even}+i\mathbf{N}\right]G^{11}+\left[\mathbf{D}_{odd}-i\mathbf{N}\right]G^{21}&=&i\mathbf{1}\nn
\left[\mathbf{D}_{odd}+i\mathbf{N}\right]G^{11}+\left[D+\mathbf{D}_{even}-i\mathbf{N}\right]G^{21}&=&0\nn
\left[D+\mathbf{D}_{even}+i\mathbf{N}\right]G^{12}+\left[\mathbf{D}_{odd}-i\mathbf{N}\right]G^{22}&=&0\nn
\left[\mathbf{D}_{odd}+i\mathbf{N}\right]G^{12}+\left[D+\mathbf{D}_{even}-i\mathbf{N}\right]G^{22}&=&-i\mathbf{1}
\tea

\bea
\mathbf{D}_{even}\left( x,y\right)=\frac12\left[\mathbf{D}\left( x,y\right)+\mathbf{D}\left( y,x\right)\right]\nn
\mathbf{D}_{odd}\left( x,y\right)=\frac12\left[\mathbf{D}\left( x,y\right)-\mathbf{D}\left( y,x\right)\right]
\tea
We obtain a more efficient representation of the dynamics by introducing new propagators

The Hadamard propagator

\begin{equation}
G_{1}=G^{21}+G^{12}\equiv <\left\{ \Phi \left(
x\right) ,\Phi \left( x^{\prime }\right) \right\} >\label{dd19r}
\end{equation}
is real and even. The
Jordan propagator

\begin{equation}
G=G^{21}-G^{12}\equiv <\left[ \Phi \left( x\right) ,\Phi
\left( x^{\prime }\right) \right] >\label{ddd19r}
\end{equation}
is imaginary and odd


The advanced and retarded propagators are the fundamental solutions for the
equations of motion for linear fluctuations in the field.

\bea
G_{ret}\left( x,x^{\prime }\right) &=&i\left[G^{11}-G^{12}\right]=
iG\left( x,x^{\prime }\right) \theta \left( t-t^{\prime
}\right)\nn
G_{adv}\left( x,x^{\prime }\right) &=&-i\left[G^{21}-G^{11}\right]=-iG\left( x,x^{\prime
}\right) \theta \left( t^{\prime }-t\right) 
\tea

All propagators may be expressed in terms of the Jordan and Hadamard ones
\begin{equation}
G^{\pm }\left( x,x^{\prime }\right) =\frac 12\left[ G_1\left( x,x^{\prime
}\right) \pm G\left( x,x^{\prime }\right) \right]
\end{equation}

\begin{equation}
G_{F,D}\left( x,x^{\prime }\right) =\frac{1}{2}\left[ G_{1}\left(
x,x^{\prime }\right) \pm G\left( x,x^{\prime }\right) \mathrm{sign}\left(
t-t^{\prime }\right) \right] 
\end{equation}
In terms of the new propagators the equations of motion are

\bea
\left[D+\mathbf{D}_{even}+\mathbf{D}_{odd}\right]G_{ret}&=&-\mathbf{1}\nn
\left[D+\mathbf{D}_{even}+\mathbf{D}_{odd}\right]G_1&=&\mathbf{N}G_{adv}
\tea


We can now begin the discussion of our subject matter. Our goal is to recast the equations for the propagators in a way suitable to discuss the last stages of the equilibration process. We shall reduce the equations to the form of a kinetic equation, the so-called Kadanoff-Baym equation. Further approximations reduce this to the Boltzmann equation. To succeed, we need a way to identify the important terms and discard the irrelevant ones. This will be provided by the so-called adiabatic expansion, under the assumption that the propagators are almost translation invariant close enough to equilibrium.

We say that a $G^{ab}\left(x.x'\right)$ is almost
translation-invariant
if, when partially Fourier transformed with respect to $u=x-x'$, the
Fourier transform is weakly dependent on the ``centroid"
variable $X=\left(x+x'\right)/2$, i.e.,

\begin{equation}
G^{ab}\left( x,x^{\prime }\right) = \int \frac{d^{d}k}{\left( 2\pi \right) ^{d}}%
\;e^{iku}G^{ab}\left( X,k\right)
\label{fast-slowb}
\end{equation}
Observe that
\bea
\int\;dy\;A\left(x,y\right)B\left(y,x'\right)=\int \frac{d^{d}k}{\left( 2\pi \right) ^{d}}%
\;e^{iku}\nn
\left\{A\left(X,k\right)B\left(X,k\right)-\frac i2\left\{A,B\right\}+\ldots\right\}
\tea
where
\be
\left\{A,B\right\}=\frac{\partial A}{\partial k}\frac{\partial B}{\partial X}-\frac{\partial A}{\partial X}\frac{\partial B}{\partial k}
\te

Expressions involving $G^{ab}\left(
X,k\right) $ may be classified according to their \emph{adiabatic order}, namely, the number of $X$ derivatives
appearing in the expression. We call this the \emph{adiabatic
expansion}. When almost
translation-invariance is verified, we may further reject all terms
above a given adiabatic order. We call such a truncation of an
adiabatic expansion an \emph{adiabatic approximation}. In other words, the adiabatic order is used as a tag
to bunch together certain terms in the equations of motion in
accordance to their derivative orders and the adiabatic approximation
determines how many of those terms are kept.

Let

\begin{equation}
i\Gamma\left( X,k\right)=i\pi\gamma\left( X,k\right)\mathrm{sign}\left(\omega\right)=\mathbf{D}_{odd}\left( X,k\right)
\end{equation}
$\omega=k^0$

\begin{equation}
R\left( X,k\right) =\left( k^{2}+m_{b}^{2}\right) -
\mathbf{D}_{even}\left( X,k\right)
\end{equation}


Then

\be 
G_{ret}\left( X,k\right)=\frac 1{R-i\Gamma}
\te

\be 
G_{adv}\left( X,k\right)=G_{ret}\left( X,-k\right)=\frac 1{R+i\Gamma}
\te

The relationship $G_{ret}=
iG\theta \left( t-t^{\prime
}\right)$ implies

\be 
G_{ret}\left(X,\left(\omega,\vec{k}\right)\right)=-\int\;\frac{d\omega'}{2\pi}\frac{G\left(X,\left(\omega',\vec{k}\right)\right)}{\omega-\omega'+i\epsilon}
\te
or else
\bea
\mathrm{Re}G_{ret}\left(X,\left(\omega,\vec{k}\right)\right)&=&-PV\int\;\frac{d\omega'}{2\pi}\frac{G\left(X,\left(\omega',\vec{k}\right)\right)}{\omega-\omega'}\nn
\mathrm{Im}G_{ret}\left( X,k\right)&=&\frac12G\left( X,k\right)
\tea
Thus

\be
G\left( X,k\right)=\frac{2\Gamma}{R^2+\Gamma^2}
\te


Under equilibrium conditions the Kubo - Martin - Schwinger theorem implies that

\be
G_{1eq}\left( X,k\right)=\mathrm{sign}\left(\omega\right)G_{eq}\left( X,k\right)\left[1+2f_{BE}\right]
\te
where $f_{BE}$ is the Bose - Einstein distribution. We generalize this to a nonequilibrium situation
by defining the \emph{density of states} ${\cal D} \left( X,k\right) $ out of
the Fourier transform of the Jordan propagator 

\begin{equation}
{\cal D} \left( X,k\right)\equiv \frac 1{2\pi}G\left( X,k\right)   \;\mathrm{sign}%
\left( \omega\right)=\frac{\gamma}{R^2+\Gamma^2}\label{cald_ch11}
\end{equation}


We now define the distribution function $f\left( X,k\right) $
through the partial Fourier transform of the Hadamard propagator

\begin{equation}
G_{1}\left( X,k\right) \equiv 2\pi {\cal D} \left( X,k\right) \;F_{1}\left(
X,k\right)  \label{hadamard}
\end{equation}

\begin{equation}
F_{1}\left( X,k\right) =1+2f\left( X,k\right)
\end{equation}



To obtain the dynamics of the distribution function $f$, we make use
of the equation involving the noise kernel. Let us call $F^{21}=\theta\left(\omega\right)+f$,  $F^{12}=\theta\left(-\omega\right)+f$,
$\Sigma_{12}=i\left(\mathbf{N}-\Gamma\right)$ and $\Sigma_{21}=i\left(\mathbf{N}+\Gamma\right)$. Then to first adiabatic order we get
\begin{equation}
A\left\{ R,F_{1}\right\} -B\left\{ \mathbf{\Gamma ,}F_{1}\right\}
=I_{col}\;\mathrm{sign}\left( k^{0}\right)   \label{ke1}
\end{equation}
where
\begin{equation}
A=\frac{\mathbf{\Gamma }^{2}}{R^{2}+\mathbf{\Gamma }^{2}}
\end{equation}
\begin{equation}
B=\frac{R\mathbf{\Gamma }}{R^{2}+\mathbf{\Gamma }^{2}}
\end{equation}
and  $I_{col}$ is the \emph{collision integral}

\begin{equation}
I_{col} =-i\left[ \Sigma_{12}F^{21}-\Sigma_{21}F^{12}\right]
\end{equation}

For weakly coupled theories, a series of approximations allow us to reduce
the  off-shell kinetic equation  to the more familiar Boltzmann kinetic equation.
We observe that in terms of the coupling constant $\lambda $ we have,
for a generic momentum $p$, $R\sim O\left( 1\right) $ while
$\mathbf{\Gamma }\sim O\left( \lambda ^{2}\right) $.

A second observation is that in general $\mathbf{\Gamma}$, which
involves the coupling constants, will be much smaller than $R$ for a
generic choice of $p$. When the coupling constants go to zero
$\mathbf{\Gamma}\to 0$, but the retarded propagator has a
well-defined asymptotic value, and the density of states becomes
$ \mathcal{D}=\delta (R)$

 In this limit the propagators are
insensitive to the behavior of the distribution function ``off shell"
(i. e., when $R\neq 0$), because the distribution function is always
multiplied by the density of states, and this is very small there.
Therefore, only ``on shell'' modes (i. e., those for which $R=0$)
really contribute to the field correlation functions. If our only
concern is to follow the evolution of the distribution function on
shell, we are allowed to replace the $A$ and $B$ coefficients in
(\ref{ke1}) by their ``on shell'' values, namely $A=1$ and $B=0.$ We
thus obtain the Kadanoff-Baym equations \cite{kad62}

\begin{equation}
\left\{ R,F_{1}\right\} =-i\;\mathrm{sign}\left( k^{0}\right) \;\left[
\Sigma_{12}F^{21}-\Sigma_{21}F^{12}\right]   \label{ke2}
\end{equation}
The
nontrivial content of the Kadanoff-Baym equations is given by the form of the collision
integral, namely, which Feynman graphs contribute to the self energies. We
recognize the structure of the collision term as the difference between a
gain and a loss term for particles moving in or out of a phase space cell
around the point $\left( X,k\right) $ per unit time. Taking $\omega >0$ for
simplicity, we see that $ \Sigma_{12}F^{21}$ is the gain
term, with $F^{21}=1+f$ accounting for stimulated emission of particles
into the cell, while the other term is the loss term, which is proportional
to the number of particles $F^{12}=f$ already there.

If we only keep the first term in the expansion, which for a $\lambda\phi^4$ theory is the setting-sun graph, we recover the Boltzmann's collision term \cite{CH88}

\end{document}